\documentclass[%
 reprint,
 amsmath,amssymb,
 aps,
]{revtex4-2}

\usepackage{graphicx}
\usepackage{dcolumn}
\usepackage{bm}
\usepackage[mathlines]{lineno}
\usepackage{xcolor}

\begin{document}

\preprint{APS/123-QED}

\title{Optomechanically Induced Transparency/Absorption in a 3D Microwave Cavity Architecture at Ambient Temperature}

\author{Sumit Kumar,$^{1}$, Matthew Kenworthy$^{1}$, Henry Ginn$^{1}$, Xavier Rojas$^{1,^*}$}
\affiliation{$^1$Department of Physics, Royal Holloway University of London, Egham, Surrey, TW20 0EX, U.K.}
\date{\today}
\thanks{xavier.rojas@rhul.ac.uk}

\begin{abstract}
Leveraging advancements in cavity optomechanics, we explore Optomechanically Induced Transparency/Absorption (OMIT/OMIA) in the microwave domain at ambient temperature. Contrary to previous works employing cryogenic temperatures, this work exploits a 3D microwave cavity architecture to observe these effects at ambient temperature, broadening the scope of possible applications. The work successfully enhances the optomechanical coupling strength, enabling observable and robust OMIT/OMIA effects, and demonstrating up to 25 dB in signal amplification and 20 dB in attenuation. Operating in the unresolved sideband regime enables tunability across a wider frequency range, enhancing the system's applicability in signal processing and sensing. The findings herein highlight the potential of optomechanical systems, presenting a simplified, cost-effective, and more feasible approach for applications at ambient temperature.
\end{abstract}

\pacs{Valid PACS appear here}

\maketitle


\section{Introduction}
Cavity optomechanics, harnessing the interaction between light and mechanical motion, has seen rapid advancements over the last two decades \cite{Aspelmeyer2014}. A particular phenomenon of interest in this domain is Optomechanically Induced Transparency (OMIT) and its counterpart, Optomechanically Induced Absorption (OMIA). These effects arise due to the strong coupling between optical/microwave and mechanical modes, leading to a modification of the cavity response \cite{Weis2010,Teufel2011,Agarwal2010,Safavi-Naeini2011,Massel2011}.

Traditional microwave optomechanics setups operate at cryogenic temperatures to suppress thermal noise, which could otherwise mask the weak optomechanical effects \cite{Palomaki2013a,Palomaki2013b,Suh2014,Wollman2015,Ockeloen-Korppi2016,Kumar2022}. However, there has been a burgeoning interest in realizing these effects at ambient temperature \cite{Faust2012,Norte2016,Pearson2020,Serra2021,Kumar2023}. Operating at ambient temperature offers several advantages such as simplifying experimental setups, cost reductions, and enhanced feasibility for various applications, especially in the realm of signal processing and sensing.

Microwave cavity optomechanics at ambient temperature presents a unique set of challenges and opportunities, especially when cooling to cryogenic temperatures is not feasible or practical. The present work leverages a 3D microwave cavity architecture to overcome some of these challenges and demonstrate the potential of OMIT/OMIA for applications in signal processing and sensing. The emphasis on ambient temperature operations facilitates experiments with a high photon number, enabling robust and clear observations of the cavity optomechanical effects.

In this work, we enhanced the optomechanical coupling strength up to $g=g_0\sqrt{n_\text{d}}\simeq 2\pi\times 95$ kHz, corresponding to cooperativity $C=4g_0^2 n_\text{d}/\kappa_\text{t}\Gamma_\text{m}\simeq30$, an important factor in achieving pronounced OMIT/OMIA effects. This coupling strength enhancement opens avenues for a range of signal processing and sensing applications. This allowed us to demonstrate up to 25 dB in signal amplification, 20 dB in signal attenuation, as well as mass sensing improvement by a factor of 5 compared to the direct measurement of mechanical resonance frequency. Our cavity optomechanical system operates in the unresolved sideband regime (i.e. bad cavity limit) allowing fine frequency tuning across the entire cavity bandwidth ($\kappa_{\rm t}/2\pi=28.5$ MHz). In the regime, where the mechanical frequency ($\Omega_\text{m}$) of our system is less than the $\kappa_{\rm t}$, the system is able to function effectively across a variety of frequencies within this bandwidth. This capability is significant because it allows for a range of operational frequencies, rather than being fixed to a single resonant frequency. Additionally, by adjusting the resonance frequency of our cavity, we can achieve finer tuning across a wider spectrum. This is achieved by altering the gap between the re-entrant post and the Si$_3$N$_4$ membrane possible by modifying the cavity design, as detailed in the studies by Floch \textit{et al.} \cite{Floch2013} and Carvalho \textit{et al.} \cite{Carvalho2016}. This feature significantly widens the range of potential applications for our design.
\newline
\begin{figure}[b]
    \centering
    \includegraphics[width=\columnwidth]{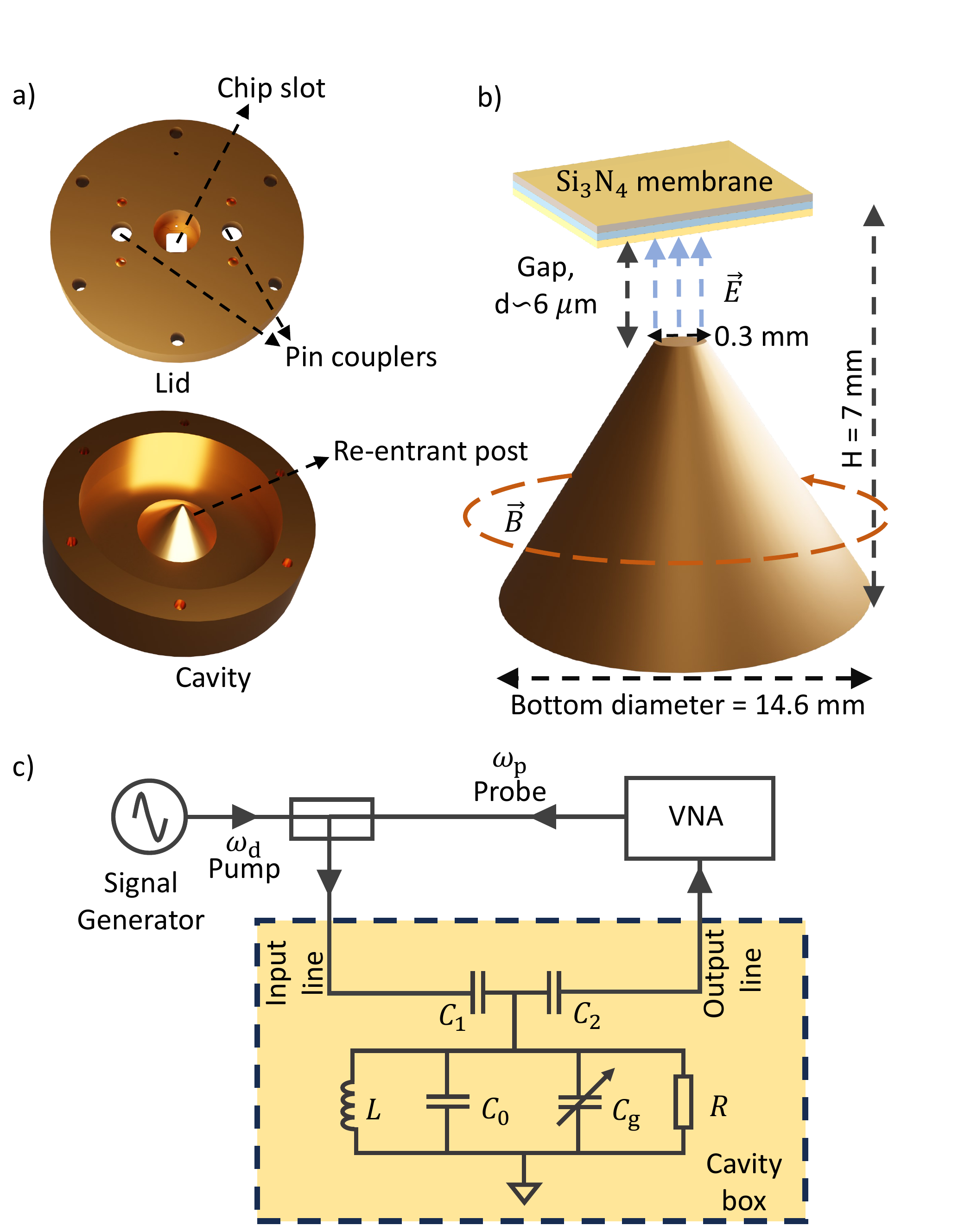}
    \caption{(a) Schematic of the microwave re-entrant cavity showing the chip window for locating the Si$_3$N$_4$ membrane chip which is mechanically clamped using a teflon screw, and the pin holes for the couplers. (b) Schematic showing the re-entrant pillar, seamlessly integrated with the Si$_3$N$_4$ membrane chip. (c) Circuit diagram showing the methodology for measuring the transmission $S_{21}(\omega)$ through the microwave cavity.}
    \label{schematic}
\end{figure}

\section{Experimental apparatus}
The core of our experiment, as detailed in Fig. \ref{schematic}, is a pre-stressed silicon nitride (Si$_3$N$_4$) membrane chip commercially purchased~\cite{Norcada}. These membranes have low mechanical loss, making them particularly suitable for our experimental setup. With dimensions of $1 \text{ mm} \times 1 \text{ mm} \times 50 \text{ nm}$, the Si$_3$N$_4$ membrane, is further plated with a 55 nm thick layer of gold deposited by thermal evaporation. The fundamental mode of this mechanical membrane has a resonance frequency $\Omega_\text{m}/{2\pi} = 160 \text{ kHz}$ and  a dissipation rate of $\Gamma_\text{m}/2\pi=45$ Hz characterized by measuring the membrane's thermomechanical motion~\cite{Kumar2022}.

Integrated into our system, the membrane interacts capacitively with a 3D microwave re-entrant cavity defined by a re-entrant post localised in the center of the cavity, in close proximity of the membrane. This design ensures a minimal gap, spanning $\sim$ 6 $\mu$m between the membrane and the post. The gap was estimated by characterizing the 3D cavity dimensions and using Finite Element Modelling (FEM) simulation. (see supplementary information). In previous work, a similar design was developed by Pate \textit{et al.} with smaller gaps in the few microns range, leading to the observation of a Casimir spring and dilution effects in microwave optomechanics \cite{Pate2020}. In the re-entrant cavity geometry, the electrical field of the fundamental mode is predominantly concentrated between the membrane and the post's top, making it possible to model this region as an effective parallel plate capacitor. The effective capacitance undergoes modulations, in response to the membrane's displacement, which can be measured through the microwave cavity transmission. The cavity resonance can be measured using the measurement setup shown~Fig. \ref{schematic}.

A parallel LCR circuit models the fundamental mode of resonance of the microwave cavity. The total capacitance of this circuit $C$ has two contribution. The capacitance $C_{\rm g}$ represents the geometric capacitance between the Si$_3$N$_4$ membrane and the pillar's top, and the capacitance $C_0$ represents the effective capacitance of the rest of the microwave cavity. The microwave cavity is coupled to the input and output transmission lines of the measurement circuit via pin antenna couplers represented in the circuit by the capacitances $C_1$ and $C_2$. The cavity is driven by a strong pump at $\omega_\text{d}=\omega_\text{c}+\Delta$ from a signal source (Rohde \& Schwarz SMA100B), where $\Delta$ represents the cavity detuning. The resonant cavity is then probed by measuring the complex transmission, corresponding to the scattering parameter $S_{21}$, using a weak probe signal at $\omega_\text{p}$ generated and detected by a vector network analyzer (Rohde \& Schwarz ZNA67).

Comprehensive insights into the design of our cavity are detailed in a previous study \cite{Kumar2023}. The experiments were conducted at ambient temperature (293 K), within a vacuum chamber that consistently maintained a pressure below $5 \times 10^{-6} \text{ mbar}$.

\section{Experimental results}
The cavity exhibits a resonance frequency of $\omega_\text{c}/2\pi=$ 4.055 GHz and a total loss rate of $\kappa_\text{t}/2\pi=$ 28.5 MHz. The external feedlines are coupled to the cavity via port 1 and port 2 with loss rates of $\kappa_1$ and $\kappa_2$ respectively, such that $\kappa_\text{ext}=\kappa_1+\kappa_2=2\pi\times 16.9$ MHz. Our cavity is therefore slightly overcoupled as $\eta=\kappa_\text{ext}/\kappa_\text{t}=0.59$.
\begin{figure}[b]
    \centering
    \includegraphics[width=\columnwidth]{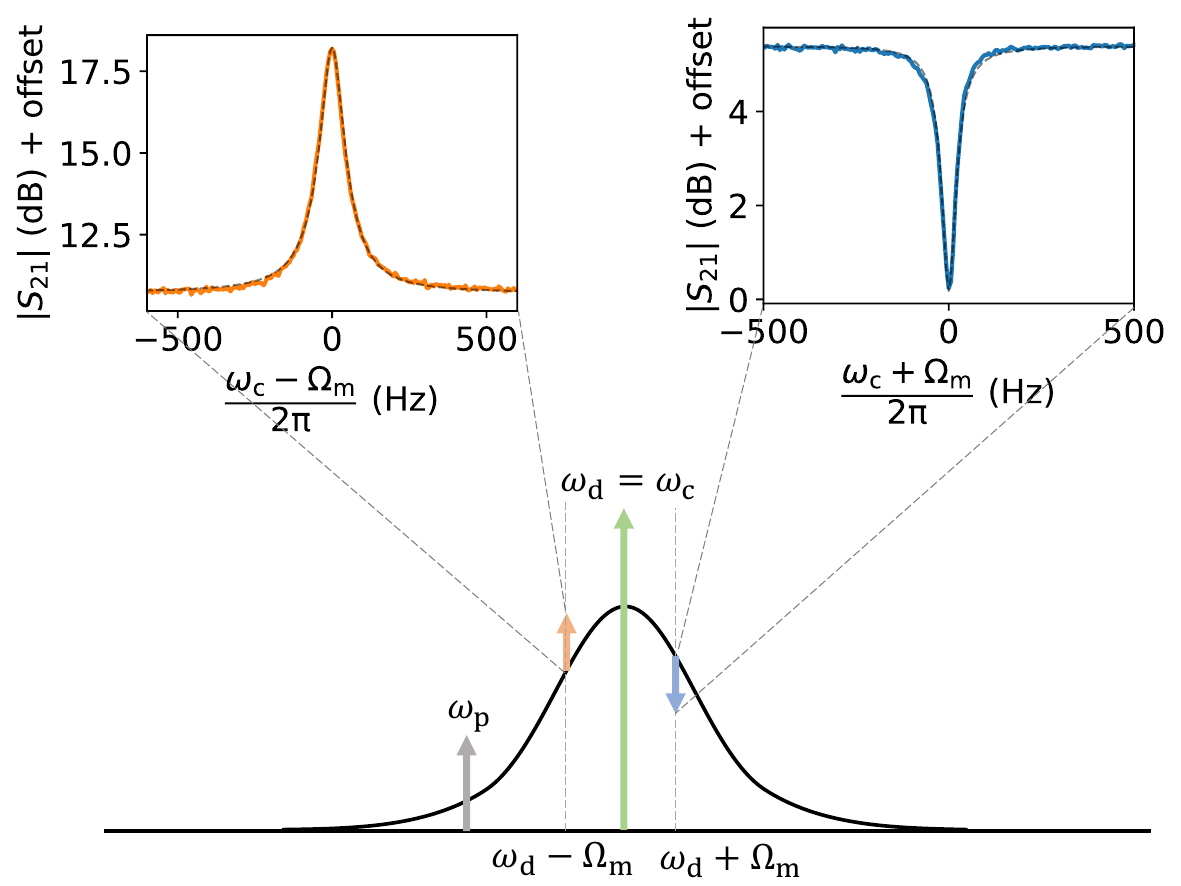}
    \caption {The cavity is driven on resonance where $\omega_\text{d}=\omega_\text{c}$, and probed with a weaker tone at $\omega_\text{p}$. The membrane's coherent drive, when $|\omega_\text{d}-\omega_\text{p}| \approx \Omega_\text{m}$, gives rise to sidebands at $\omega_\text{d} \pm \Omega_\text{m}$ with opposite phases. These sidebands interfere constructively (amplification) or destructively (absorption) with the probe signal, at $\omega_p=\omega_\text{d}\pm\Omega_\text{m}$. The inset magnifies a typical amplification (orange curve) and absorption (blue curve) for an arbitrary pump power.}
    \label{figure2}
\end{figure}

The cavity is driven by a strong pump at $\omega_\text{d}=\omega_\text{c}+\Delta$ and probed using a weaker probe tone at $\omega_\text{p}$. The beating of $\omega_\text{d}$ and $\omega_\text{p}$ results in a radiation pressure force on the membrane at the difference frequency $\Omega=\omega_\text{p}-\omega_\text{d}$. This leads to coherent drive of the membrane when $||\Omega|-\Omega_\text{m}|<<\Gamma_\text{m}$, where $\Gamma_\text{m}$ is the mechanical linewidth. The signal at $\omega_\text{d}$ is modulated due to driven mechanics, creating sidebands at $\omega_\text{d}\pm \Omega_\text{m}$ as shown in Fig. \ref{figure2}. In this study, we performed the measurements at zero detuning ($\Delta=0$) that is $\omega_\text{d}=\omega_\text{c}$, for which we do not expect dynamical backaction effects (cooling/heating) on the mechanical system.

\begin{figure}
	\centering
	\includegraphics[width=\columnwidth]{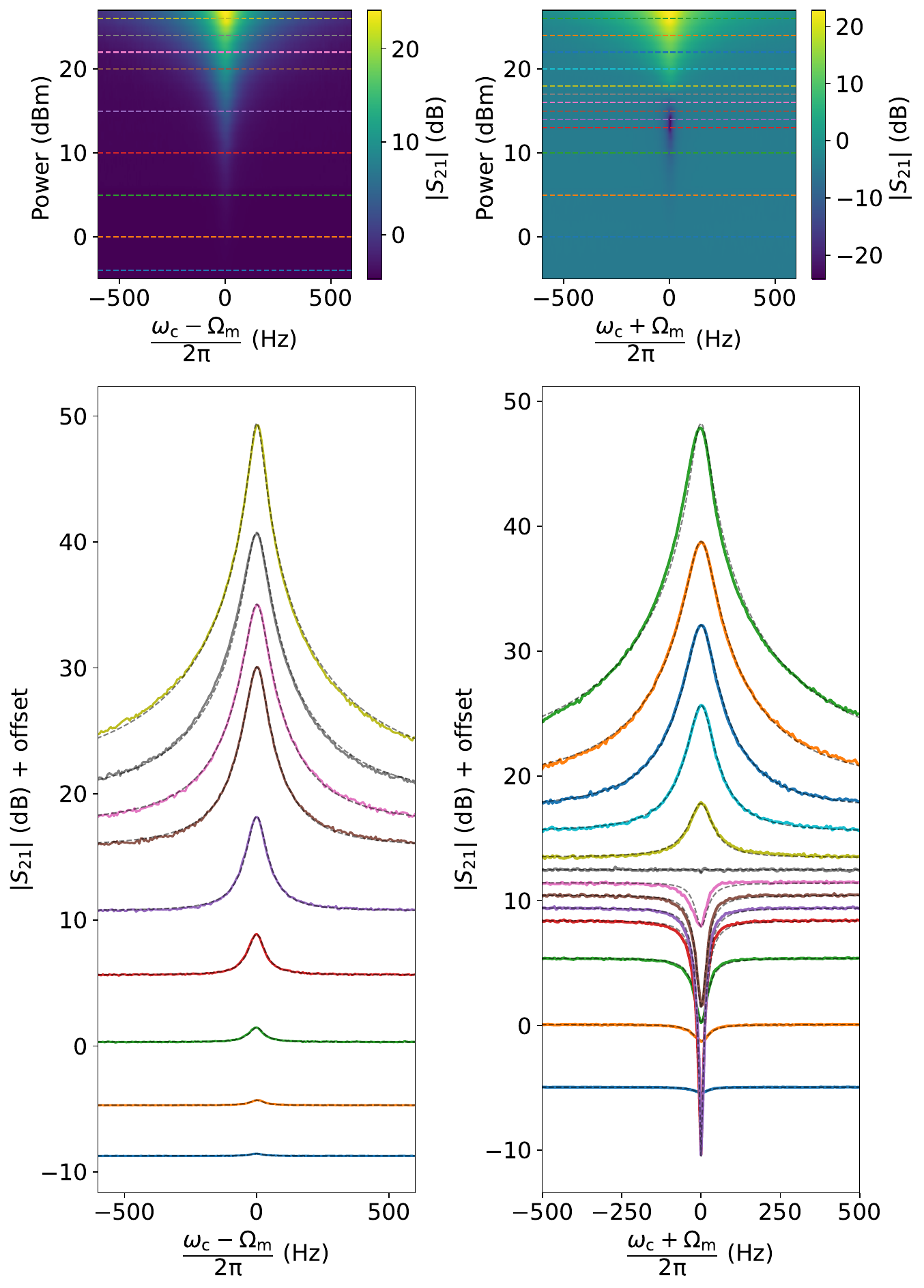}
	\caption{\label{figure3} (left) Colormap showing $S_{21}$ for different pump powers (-5 dBm to 27 dBm) at $\omega_\text{c}-\Omega_\text{m}$. We see continuous amplification with increasing pump power. The colored linecuts correspond to $S_{21}$ at different pump powers. The y-axis is offset for different curves for clarity. The fits are shown in dashed grey. (right) Colormap showing $S_{21}$ for different pump powers (-5 dBm to 27 dBm) at $\omega_\text{c}+\Omega_\text{m}$. We see absorption from -5 dBm to 15 dBm and then amplification for powers above 16 dBm. The colored linecuts corresponds to $S_{21}$ at different pump powers. The y-axis is offset for different curves for clarity. The fits are shown in dashed grey.  }
\end{figure}

The red sideband at $\omega_\text{c}-\Omega_\text{m}$ interferes constructively with the probe signal, leading to amplification of the output signal. In this regime the system can be used as an amplifier for the probe signal. The amplification gain is tunable with the pump power. We measured amplification gain up to 25 dB, which was limited by the signal generator maximum output power. This effect is shown in Fig. \ref{figure3}.  

Below an input pump power threshold of 16 dBm, corresponding to a cavity photons $n_d\sim4.5\times10^{13}$ photons in this system, the red-detuned sideband at $\omega_\text{c}+\Omega_\text{m}$ is in opposite phase with the probe signal, and interferes destructively with it, leading to signal absorption. In this regime, the system can be used as a notch filter for the probe signal. At 14 dBm of input power corresponding to cavity photons $n_d\sim2.8\times10^{13}$, we obtain a signal reduction of up to 20 dB on a narrow frequency range of $\Delta\omega/2\pi=8$ Hz, which is 5 times smaller than the mechanical linewidth ($\Gamma_{\rm m}/2\pi=45$ Hz). This effect is shown in Fig. \ref{figure3}. This power threshold is associated with a cooperativity approaching unity ($C\sim1$).
In this regime, the cavity optomechanical system can be used as a notch filter of high selectivity and tunability in the microwave frequency range. This narrow notch filter could also be used for sensing purposes, allowing, for instance, a sensitive mass detection of 50 attogram deposited on the mechanical membrane. This corresponds to being able to detect a mechanical frequency shift of the order of half the notch filter linewidth ($\Delta\omega/2 = 2\pi\times4$ Hz), which is 5 times smaller than the half bare mechanical linewidth. As demonstrated, leveraging cavity optomechanical absorption for mass detection can substantially enhance sensitivity compared to solely relying on the mechanical resonator detection alone. Theoretically, as the cooperativity, $C$, approaches 1, one can achieve unlimited signal absorption over an infinitely narrow linewidth. In principle, this would enable the detection of incredibly minute masses. However, the ultimate limit to this detection capability is set by frequency fluctuations of the mechanical resonator. Provided that one could reach such limit of $C$ approaching 1, this system would be an ideal probe to study frequency fluctuations in mechanical resonators, which are not yet totally understood \cite{maillet2018}. 

Above the input power threshold of 16 dBm, the blue sideband at $\omega_\text{c}+\Omega_\text{m}$ interfere constructively with the probe signal leading to signal amplification as for the blue-detuned sideband.

We measured the transmission of the probe signal through the cavity $S_{21}$ in the vicinity of $\omega_\text{c} \pm \Omega_\text{m}$ for different pump powers (-5 dBm to 27 dBm) applied at the signal generator output, hence spanning cavity photon numbers over three orders of magnitude ($n_{\rm d} \simeq 10^{12}$ to $10^{14}$ photons). Fig. \ref{figure3} shows a colormap of $S_{21}$, highlighting the amplification and absorption of the probe signal after interfering with the motional sidebands. The blue-detuned sideband at $\omega_\text{c}-\Omega_\text{m}$ is continuously amplified with increasing pump power, while absorption was observed at the red-detuned sideband $\omega_\text{c}+\Omega_\text{m}$ for low powers. At higher pump powers (16 dBm and above), however, we observed amplification at the red-detuned sideband $\omega_\text{c}+\Omega_\text{m}$. Similar results have been obtained in previous microwave optomechanical setups operated at cryogenic temperatures~\cite{Massel2011,cohen2020}. This work shows that these effects can also be observed at ambient temperature. Fig.~\ref{figure3} also shows line cuts of $S_{21}$ along the colormap for different powers corresponding to different cavity photons numbers $n_\text{d} = 2\kappa_\text{ext}P_\text{in}/ \hbar \omega_\text{c}\kappa_\text{t}^2$ where $P_\text{in}$ is the power applied at the cavity input, taking into account 3.3 dB of attenuation from the cables. Theoretically, the cavity transmission $S_{21}$ is given as,
\begin{equation}\label{Eq_OMIT}
	S_{21}= \dfrac{\kappa_\text{ext}[1+i\chi]}{\kappa_\text{t}-2i(\omega_\text{p}-\omega_\text{c})}{\text{,}}
\end{equation}
where, 
\begin{equation}
	\chi= \dfrac{4\Omega_\text{m}g_0^2 n_\text{d}}{[\Omega_\text{m}^2-(\omega_\text{p}-\omega_\text{c})^2-i\Gamma_\text{m}(\omega_\text{p}-\omega_\text{c})][\kappa_\text{t}-2i(\omega_\text{p}-\omega_\text{c})]}{\text{,}}
\end{equation}
 and $g_0$ is the single photon optomechanical coupling strength. See supplementary information for a derivation of the transmission. Using the above equations, we fitted our data at different powers and extracted the single photon optomechanical coupling strength $g_0/2\pi=$ 4 mHz. This represents an improvement of an order of magnitude compare to our previous architecture \cite{Kumar2023}. The other fitting parameters were the cavity linewidth $\kappa_\text{t}$ and the mechanical linewidth $\Gamma_\text{m}$, which have a slight dependence on the photon number $n_\text{d}$ inside the cavity. The ratio $\kappa_\text{t}/\Omega_\text{m}=178$ indicates that our system is in the unresolved sideband regime (or bad cavity limit).

The microwave resonance frequency $\omega_\text{c}$ drifts over time with changes in the ambient temperature due to temperature dependent cavity parameters. Consequently, $\omega_\text{c}$ was monitored to adjust the frequency of the pump signal maintaining the zero detuning condition ($\Delta=0$). A small power-dependent cavity damping was also observed, which suggests that non-linear effects come into play at higher driving powers. Since we obtained an excellent agreement fitting the cavity response at each power with a standard harmonic oscillator equation, we simply defined a power-dependent linewidth $\kappa_{\rm t}$.

At high power, we also observed a frequency shift of the mechanical resonance frequency, which can be attributed to changes in the membrane's tension caused by thermal expansion. This spring softening of the membrane at the highest power is due to classical heating of the membrane, which was described in our previous work \cite{Kumar2023}. We would also like to mention that the frequency shift is not due to  This drift of the mechanical resonance frequency, is also monitored to adjust the probe signal frequency.

\begin{figure}
	\centering
	\includegraphics[width=\columnwidth]{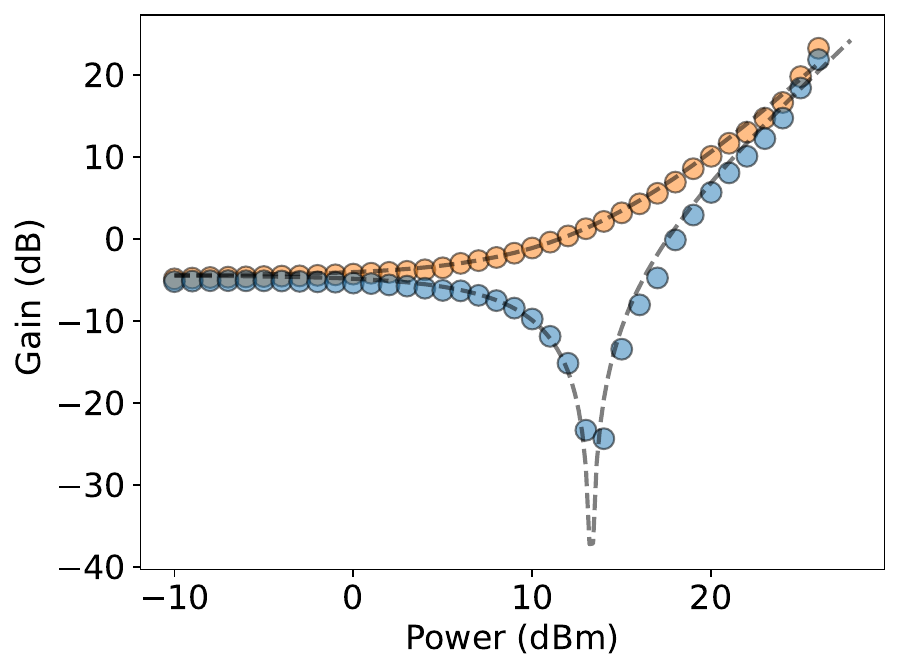}
	\caption{\label{figure4} Plot showing the gain/attenuation originating from the cavity optomechanical system and its generated sidebands interfering with the probe signal. The left sideband at $\omega_\text{c}-\Omega_\text{m}$ (orange datapoints), and the right sideband at $\omega_\text{c}+\Omega_\text{m}$ (blue datapoints).}
\end{figure}

The maximum gain/attenuation for each power used in this experiment is plotted on Fig. \ref{figure4}. Eq. \ref{Eq_OMIT} fits well at most power, indicative of a good agreement with the theoretical description of OMIT/OMIA.

The noise due to thermal fluctuations can significantly impact the signal at cavity resonance. The noise stems from thermal occupancy of both the cavity and mechanical resonator at ambient temperature. The thermal fluctuations of the Si$_3$N$_4$ membrane introduce more noise compared to the cavity, primarily because cavity resonance frequency is much larger than mechanical resonance frequency i.e. $\omega_\text{c}>>\Omega_\text{m}$. This disparity in frequencies accentuates the influence of the membrane’s thermal fluctuations, making them a dominant source of noise in the system. The thermal fluctuations experienced by the membrane are influenced by its effective temperature, which is  influenced by a several factors including the temperature of the surrounding bath, the phase noise of the pump signal, and classical heating resulting from microwave absorption. To fully characterise the performances of these devices as amplifiers/filters, further studies of the noise should be carried.

\section{Conclusion}
In this work, we have detailed the experimental setup and results for studying optomechanical interactions in a microwave re-entrant cavity with a Si$_3$N$_4$ membrane. Our results clearly demonstrate OMIT and OMIA effects of large amplitude, with a good agreement between the experimental observations and theoretical predictions. The precise control and measurement of these effects highlight the potential of our system for further optomechanical studies and applications in signal processing and sensing.

Despite the excellent agreement between experiment and theory, our observations at high power levels indicate a lack of temperature stability of our cavity optomechanical system. Improving temperature control in the system and increasing the strength of coupling would help reduce the impact of parasitic effects. Also, the operating bandwidth of our system could be broadened by tuning $\omega_\text{c}$, achieved by adjusting the gap between the membrane and the top of the conical post. This adjustment allows for enhanced flexibility and a wider range of detectable frequencies. Further research in these directions could provide additional insights and enable the enhancement of optomechanical interactions for advanced applications.

\section*{Data availability}
The data that support the findings of this study are openly available in zenodo at https://doi.org/10.5281/zenodo.10560931, Ref.~\cite{data-zenodo}

\section*{Supplementary material}
The supplementary material presents our estimation of the reentrant cavity gap based on FEM simulations, and the derivation of the cavity transmission $S_{21}$ obtained from the equations of motion, when the cavity is driven by a control and a probe signal.

\section*{acknowledgment}
We thank A. Armour for helpful discussions, and C. Guisca for his help on the surface characterization of our 3D cavities. This research was supported by the Royal Society (UF150140, RGF\textbackslash EA\textbackslash 180099, RGF\textbackslash R1\textbackslash 180059, RGF\textbackslash EA\textbackslash 201047, RPG\textbackslash 2016\textbackslash 186, URF\textbackslash R\textbackslash 211009, and RF \textbackslash ERE\textbackslash 210198), the EPSRC (EP/R04533X/1), and the STFC (ST/T005998/1). Measurements were made at the London Low Temperature Laboratory, supported by technical staff, in particular R. Elsom, I. Higgs, P. Bamford, and H. Sandhu. For the purpose of open access, the author has applied a Creative Commons Attribution (CC BY) license (where permitted by UKRI, “Open Government Licence” or “Creative Commons Attribution No-derivatives (CC BY-ND) license” may be stated instead) to any Author Accepted Manuscript version arising.

\bibliography{aipsamp}

\section*{Supplementary 1: Estimating the gap between re-entrant post and the membrane}\label{simulation}
\begin{figure}[h]
			\centering
			\includegraphics[width=\columnwidth]{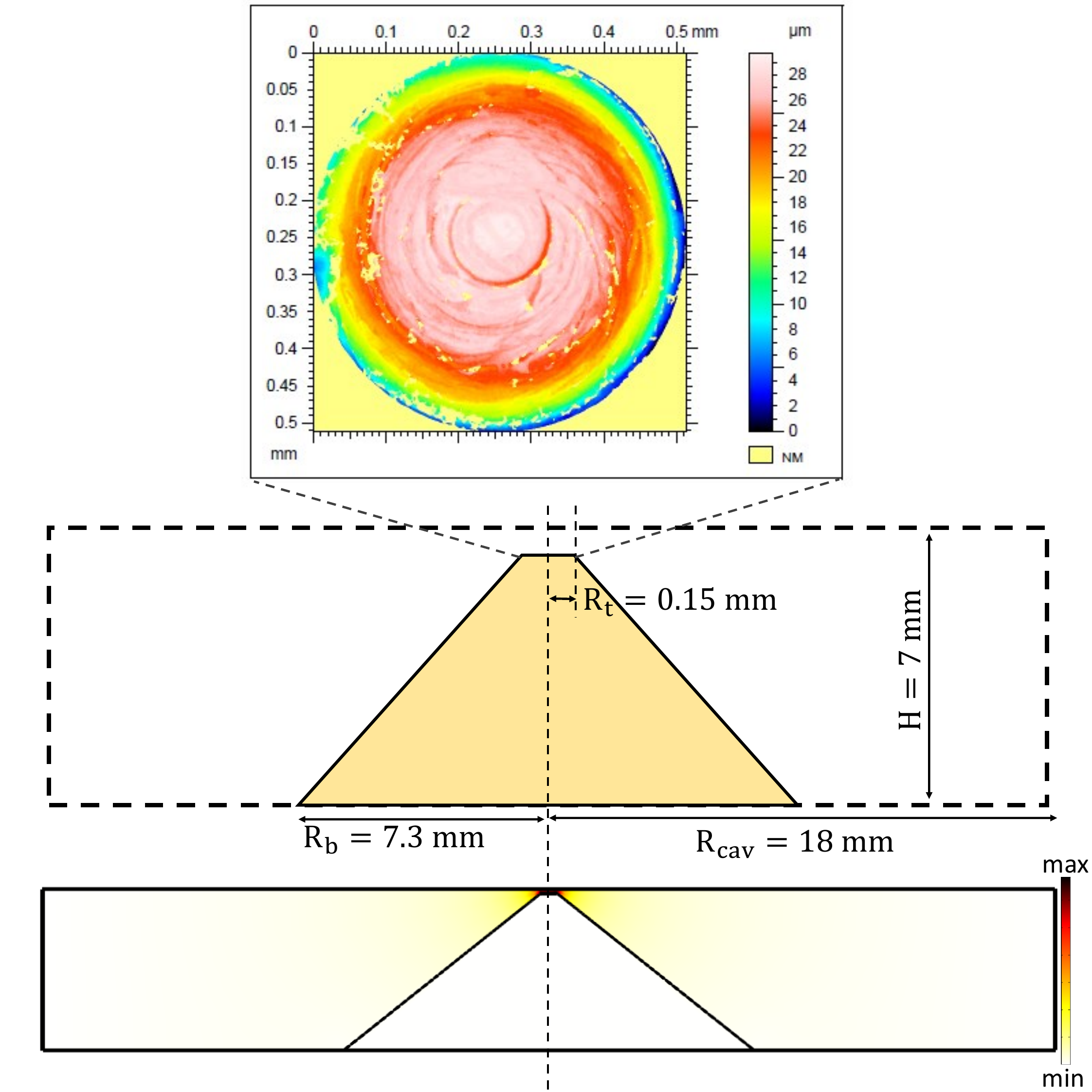}
			\caption{\label{figure5} (top) Figure showing the surface profile of the top of the re-entrant post, obtained using an optical interferometer, with the dimensions of the re-entrant cavity shown below. (bottom) Electric field profile of the simulated re-entrant cavity mode at 4.055 GHz obtained using an FEM simulation (COMSOL).}
		\end{figure}

		The dimensions of the top of the re-entrant post are determined using optical interferometery. The estimated effective radius of the top of the re-entrant post is $\rm R_t=0.15$ mm. The resonance frequency of the 3D re-entrant cavity is predominantly influenced by the separation between the top of the re-entrant post and the membrane. We conducted Finite Element Method (FEM) simulations using COMSOL, depicted in Fig. \ref{figure5}, to determine the microwave resonance frequency. A resonance frequency of 4.055 GHz, obtained from the simulation, corresponds to the measured resonance frequency given a 6 $\mu$m gap between the post top and the silicon nitride membrane.
		
\section*{Supplementary 2: Derivation of the equations of motion}\label{derivation}
		We'll begin by formulating the classic optomechanical equation of motion for both the mechanical resonator and the optical cavity given by,
		\begin{equation}{\label{A1}}
			\ddot{x}=-\Omega_\text{m}^2 x - \Gamma_\text{m}\dot{x}+\dfrac{(F_\text{BA}+F_\text{th})}{m_\text{eff}},
		\end{equation}
		and
		\begin{equation}{\label{A2}}
			\dot{\alpha}=[-\frac{\kappa_\text{t}}{2}+i\Delta]\alpha+iGx \alpha+\sum_\text{k=1,2} \sqrt{\kappa_\text{k}}\alpha_\text{in,k}.
		\end{equation}
		The motion of the mechanical resonator can be described by Eq. \ref{A1} where $m_\text{eff}$ represents the effective mass, $x$, $\dot{x}$, and $\ddot{x}$ are respectively the membrane's displacement, velocity and acceleration, $F_\text{BA}=\hbar G |\alpha| ^2 $ is the backaction force, and $F_\text{th}$ is the thermal force. $\alpha$ is the intracavity field and $G=\partial \omega_\text{c}/\partial{x}$ is the frequency pull factor. The equation of motion for the intracavity field $\alpha$ in the frame rotating with pump frequency $\omega_\text{d}$ is expressed by Eq. \ref{A2} where $\kappa_\text{t}$ is the total cavity decay rate, $\kappa_\text{k=1,2}$ is the coupling rate of cavity to left (1) and right (2) ports respectively, $\alpha_\text{in,k}$ is the input field, and $\Delta=\omega_\text{d}-\omega_\text{c}$. In the context of OMIT/OMIA, random noise inputs to the mechanical resonator and the cavity are neglected. Eq. \ref{A1} and Eq. \ref{A2} are reframed as,
		\begin{equation}{\label{A3}}
			\ddot{x}=-\Omega_\text{m}^2 x - \Gamma_\text{m}\dot{x}+\dfrac{\hbar G |\alpha| ^2}{m_\text{eff}},
		\end{equation}
		and
		\begin{equation}{\label{A4}}
			\dot{\alpha}=[-\frac{\kappa_\text{t}}{2}+i\Delta]\alpha+iGx \alpha+ \sqrt{\kappa_\text{1}}\alpha_\text{in}+\sqrt{\kappa_\text{1}}s_\text{p},
		\end{equation}
		where, $s_\text{p}=s_0 e^{-i\Omega t} $ is the probe field, $\Omega=\omega_\text{p}-\omega_\text{d}$, and $\alpha_\text{in}$ is the input pump field. We will solve Eq. \ref{A3} and Eq. \ref{A4} by linearizing them around their steady-state solutions, $\bar{\alpha}$ and $\bar{x}$, where $x=\bar{x}+\delta x$ and $\alpha=\bar{\alpha}+\delta \alpha$. The equations of motion, once linearized, can be expressed as follows,
		\begin{equation}{\label{A5}}
			\delta \ddot{x}=-\Omega_\text{m}^2 \delta x - \Gamma_\text{m}\delta\dot{ x}+\dfrac{\hbar G \bar{\alpha}(\delta \alpha+\delta \alpha^*)}{m_\text{eff}},
		\end{equation}
		and
		\begin{equation}{\label{A6}}
			\delta \dot{\alpha}=[-\frac{\kappa_\text{t}}{2}+i\Delta]\delta\alpha+iG \bar{\alpha}\delta x+ \sqrt{\kappa_\text{1}}s_\text{p}.
		\end{equation}
		To solve Eq. \ref{A5} and Eq. \ref{A6} we use the following ansatz,
		\begin{equation}{\label{A7}}
			\delta \alpha= A_1e^{-i\Omega t}+A_2e^{i\Omega t},
		\end{equation}
		\begin{equation}{\label{A8}}
			\delta \alpha^*= A_1^*e^{i\Omega t}+A_2^*e^{-i\Omega t},
		\end{equation}
		\begin{equation}{\label{A9}}
			\delta x= Xe^{-i\Omega t}+X^*e^{i\Omega t}.
		\end{equation}
		Using above equations and only taking terms with $e^{-i\Omega t}$ coefficients, Eq. \ref{A5} simplifies to
		\begin{equation}{\label{A10}}
			X=\left[\dfrac{\hbar G \bar{\alpha}(A_1+A_2^*)}{m_\text{eff}}\right]\chi_\text{m}(\Omega),
		\end{equation}
		where
		\begin{equation}{\label{A11}}
			\chi_\text{m}(\Omega)=\dfrac{1}{m_\text{eff}\left[\Omega_\text{m}^2-\Omega^2-i\Gamma_\text{m}\Omega\right]}.
		\end{equation}
		Similarly separating the terms with $e^{-i\Omega t}$ and $e^{i\Omega t}$  coefficients from Eq. \ref{A6} we get,
		\begin{equation}{\label{A12}}
			A_1=\left[iG\bar{\alpha}X+\sqrt{\kappa_\text{1}}s_\text{p}\right]\chi_\text{a}(\Omega),
		\end{equation}
		and
		\begin{equation}{\label{A13}}
			A_2=\left[iG\bar{\alpha}X^*\right]\chi_\text{b}(\Omega),
		\end{equation}
		where,
		\begin{equation}{\label{A14}}
			\chi_\text{a}(\Omega)=\dfrac{1}{\frac{\kappa_{\rm t}}{2}-i(\Delta+\Omega)}, \chi_\text{b}(\Omega)=\dfrac{1}{\frac{\kappa_{\rm t}}{2}+i(\Omega-\Delta)},
		\end{equation}
		and
		\begin{equation}{\label{A15}}
			\chi_\text{c}(\Omega)=\chi^*_\text{b}(\Omega)=\dfrac{1}{\frac{\kappa_{\rm t}}{2}-i(-\Delta+\Omega)}.
		\end{equation}
		Now putting the expression of  $A_2^*$ in Eq. \ref{A10} we get,
		\begin{equation}{\label{A16}}
			X= \dfrac{A_1}{\dfrac{1}{\chi_\text{m}(\Omega)\hbar G\bar{\alpha}}+iG\bar{\alpha}\chi_\text{c}(\Omega)}.
		\end{equation}
		We now continue by putting the above expression in Eq. \ref{A12} and solve for $A_1$ giving us,
		\begin{equation}{\label{A17}}
			A_1=\sqrt{\kappa_\text{l}}s_\text{p}\left[\dfrac{1+i\chi }{\dfrac{1}{\chi_\text{a}(\Omega)}+i\chi [\dfrac{1}{\chi_\text{a}(\Omega)}-\dfrac{1}{\chi_\text{c}(\Omega)}]}\right],
		\end{equation}
		where,
		\begin{equation}{\label{A18}}
			\chi=\hbar G^2 \bar{\alpha}^2 \chi_\text{c}(\Omega)  \chi_\text{m}(\Omega).
		\end{equation}
		In our case, we have $\omega_\text{d}=\omega_\text{c}$, Eq. \ref{A18} and thus Eq. \ref{A17} simplifies to
		\begin{equation}{\label{A19}}
			\chi=
			\dfrac{4\Omega_\text{m}g_0^2 n_\text{d}}{[\Omega_\text{m}^2-(\omega_\text{p}-\omega_\text{c})^2-i\Gamma_\text{m}(\omega_\text{p}-\omega_\text{c})][\kappa_\text{t}-2i(\omega_\text{p}-\omega_\text{c})]}{\text{,}}
		\end{equation}
		and 
		\begin{equation}{\label{A20}}
			A_1=\dfrac{2\sqrt{\kappa_\text{1}}s_\text{p}[1+i\chi]}{\kappa_\text{t}-2i(\omega_\text{p}-\omega_\text{c})}.
		\end{equation}
		The scattering parameter can further be written as,
		\begin{equation}{\label{A21}}
			S_{21}=\dfrac{\sqrt{\kappa_\text{2}}}{s_\text{p}}\dfrac{2\sqrt{\kappa_\text{l}}s_\text{p}[1+i\chi]}{\kappa_\text{t}-2i(\omega_\text{p}-\omega_\text{c})}.
		\end{equation}
		Since $\kappa_\text{1}=\kappa_\text{2}=\kappa_\text{ext}/2$, above equation simplifies to 
		
		\begin{equation}{\label{A22}}
			S_{21}=\dfrac{\sqrt{\kappa_\text{ext}}[1+i\chi]}{\kappa_\text{t}-2i(\omega_\text{p}-\omega_\text{c})}.
		\end{equation}

\end{document}